\documentclass [twocolumn, aps, showpacs] {revtex4}
\usepackage{graphicx}
\usepackage{amsmath}
\usepackage{amssymb}
\usepackage{epsf}
\newlength{\upit}\upit=0.1truein

\newcommand{\ltappr}{{{\lower4pt\hbox{$<$} } \atop \widetilde{ \ \ \ }}}
\newlength{\bxwidth}\bxwidth=1.5 truein

\def\be{\begin{equation}}
\def\ee{\end{equation}}
\def\eqa{\begin{eqnarray}}
\def\eea{\end{eqnarray}}

\newlength{\figwidth}
\newlength{\shift}
\newcommand \bea {\begin{eqnarray} }

\begin{document}
\title {Quantum Algorithm to Solve Satisfiability Problems}

\author {Wenjin Mao}
\affiliation{ Department of Physics and Astronomy, Stony
Brook University, SUNY, Stony Brook, NY 11794, U.S.A.}
\date{\today}
\begin{abstract}
A new quantum algorithm is proposed to solve Satisfiability(SAT) problems by taking advantage of non-unitary transformation in ground state quantum computer. The energy gap scale of the ground state quantum computer is analyzed for 3-bit Exact Cover problems. The time cost of this algorithm on general SAT problems is discussed.
\end{abstract}
\maketitle

Quantum computer has been expected to outperform its classical counterpart in some classically difficult problems. For example, the well-known Shor's factoring algorithm\cite{Shor}  and Grover's algorithm\cite{Grover} accelerate exponentially and quadratically compared with classical algorithms.
It is  a challenge to find whether quantum computer outperforms its classical counterpart on other classically intractable problems\cite{Farhi1, Hogg}, which cannot be solved classically in polynomial time of $N$, the number of input bits. Especially interesting are  the NP-complete problems\cite{NP-Complete}, which include thousands of problems, such as the Traveling Salesman problem\cite{EC} and some satisfiability (SAT) problems.   All NP-complete problems can be transformed into each other by polynomial steps. If one of the NP-complete problems can be solved in polynomial time by an algorithm in the worst case, then all NP-complete problems can be solved in polynomial time. However, it is widely believed that such a classical algorithm doesn't exist. 


In this paper we explore the idea of ground state quantum computer (GSQC)\cite{Mizel1, Mizel2, Mizel3, ours}, and propose a new algorithm to solve SAT problems. 
A $K$-SAT problem deals with $N$ binary variables submitted to $M$ clauses with each clause $C_i$  involving $K$ bits, and the task is to find $N$-bit states satisfying all clauses.
 When $K\ge 3$, $K$-SAT is NP-Complete, and some instances become
classically intractable when the parameter $\alpha=M/N$, as $M,\ N\rightarrow\infty$, is close to threshold $\alpha_c(K)$\cite{3SAT, SAT, NP}. 


 A standard computer is characterized by time dependent state as:
$
|\psi(t_i)\rangle=U_i|\psi(t_{i-1})\rangle,
$
where $t_i$ denotes instance of  the $i$-th step, and $U_i$ represents for unitary transformation. For GSQC, the time sequence is mimicked by  the space distribution of the ground state wavefunction $|\psi_0\rangle$. 

As proposed by Mizel et.al.\cite{Mizel1}, a single qubit may be a column of  quantum dots with multiple rows, and each row contains a pair of quantum dots. State $|0\rangle$ or $|1\rangle$ is represented by finding electron in one of the two dots. GSQC is made up by circuit of multiple interacting qubits, whose ground state is determined by the summation of single qubit unitary transformation Hamiltonian $h(U_j)$, two-qubit interacting Hamiltonian $h(CNOT)$, boost Hamiltonian $h(B,\lambda)$ and projection Hamiltonian $h(|\gamma\rangle,\lambda)$. The energy gap between the ground state and the first excited state determines the efficiency of GSQC\cite{ours}.

The single qubit unitary transformation Hamiltonian has the form:
\eqa
h^j(U_j)=\epsilon\left[ 
C^{\dagger}_{j-1}C_{j-1}+C^{\dagger}_{j}C_{j} -\left(C^{\dagger}_{j}U_jC_{j-1}+h.c.\right)\right],
\eea
where  $\epsilon$ defines the energy scale of all Hamiltonians, $C^{\dagger}_j=\left[c^{\dagger}_{j,0}\ c^{\dagger}_{j,1}\right]$, $c_{j,0}^{\dagger}$ is the electron creation operator on row $j$ at position $0$, and $U_j$ is two dimension matrix representing for unitary transformation from row $j-1$ to row $j$. The boost  Hamiltonian is:
\eqa
h^j(B,\lambda)=\epsilon\left[
C^{\dagger}_{j-1}C_{j-1}+\frac{1}{\lambda^2}C^{\dagger}_{j}C_{j}
-\frac{1}{\lambda}\left(C^{\dagger}_{j}C_{j-1}+h.c.\right)\right],
\eea
which amplifies the $j$th row wavefunction amplitude by large number $\lambda$ compared with $(j-1)$th row in $|\psi_0\rangle$. The projection Hamiltonian is
\eqa
h^j\left(|\gamma\rangle,\lambda\right)=\epsilon\left[
c^{\dagger}_{j-1,\gamma}c_{j-1,\gamma}+\frac{1}{\lambda^2}c^{\dagger}_{j,\gamma}c_{j,\gamma} -\frac{1}{\lambda}\left(c^{\dagger}_{j,\gamma}c_{j-1,\gamma}+h.c.\right)\right], 
\eea
where $|\gamma\rangle$ represents for state to be projected to on row $j$ and to be amplified by $\lambda$. The interaction between qubit $\alpha$ and $\beta$ can be represented by $h(CNOT)$:
\eqa
h^j_{\alpha,\beta}(CNOT)&=&
\epsilon C^{\dagger}_{\alpha,j-1} C_{\alpha,j-1} C^{\dagger}_{\beta,j} C_{\beta,j}\nonumber\\
& &+h^j_{\alpha}(I)C^{\dagger}_{\beta,j-1} C_{\beta,j-1}
+ c^{\dagger}_{\alpha,j,0} c_{\alpha,j,0}h^j_{\beta}(I)\nonumber\\
& &
+c^{\dagger}_{\alpha,j,1} c_{\alpha,j,1}h^j_{\beta}(N). \label{hCNOT}
\eea
where for $c^{\dagger}_{a,b,c}$, its subscription $a$ represents for qubit $a$, $b$ for the number of row, $c$ for the state $|c\rangle$. 
 All above mentioned  Hamiltonians are positive semidefinite, and are the same as those in \cite{Mizel1, Mizel2, Mizel3}. Only pairwise interaction is considered for interacting Hamiltonians.

 The input states are determined by the boundary conditions applied upon the first rows of all qubits, $h^0=E(I+\sum_i a_i\sigma_i)$, with $\sigma_i$ being Pauli matrix, $\sum_ia_i^2=1$ and $E$ being large compared with $\epsilon$\cite{ours}.

To implement any algorithm, on final row of each qubit boost or projection Hamiltonian is applied so that $|\psi_0\rangle$ concentrates on the position corresponding to the final step in standard paradigm,  hence measurement on GSQC can read out desired information with appreciable probability.

\begin{figure}
\begin{center}
\leavevmode
\hbox{\epsfxsize=2cm \epsffile{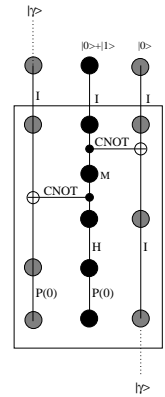}}
\end{center}
\caption{\small Design for "teleportation box". The circuit inside box is similar to the teleportation circuit in \cite{Mizel3}. Label $I$ represents for identity transformation Hamiltonian $h(I)$, $H$ for Hadamard transformation Hamiltonian $h(H)$ and $P(0)$ for projection Hamiltonian $h(|0\rangle,\lambda)$. }
\label{teleport}
\end{figure}

 As shown in \cite{ours} GSQC circuit may have exponentially small energy gap depending on detail of circuit, and assembling GSQC circuit directly following algorithm for standard paradigm, such as quantum Fourier transform, leads to exponentially small energy gap. In order to avoid small gap, teleportation box, as shown in Fig.(\ref{teleport}), is introduced on each qubit between two control Hamiltonians. The teleportation boxes make all qubits short (the longest qubit has length 8), on the other hand, for arbitrary GSQC circuit they make the energy gap only polynomially small $\Delta\propto \epsilon/\lambda^8$ if all boost and projection Hamiltonians have the same $\lambda$ value. To determine magnitude of $\lambda$, one only needs to count the total number of qubits in the circuit, which is proportional to the number of control operation in an algorithm, say $N^k$, then the probability of finding all electrons on final rows is $P\approx (1-C/\lambda^2)^{N^k}$ with $C$ being 8, the maximum length of qubit. In order to have appreciable $P$, we set $\lambda\approx N^{k/2}$, hence $\Delta\propto \epsilon/N^{4k}$. The details can be found in \cite{ours}.

While a time-dependent standard quantum computer makes unitary transformation from one instance to the next, GSQC may have non-unitary transformation from one row to the next, such as the boost Hamiltonian $h(B,\lambda)$ and projection Hamiltonian $h(|\gamma\rangle, \lambda)$. Especially the projection Hamiltonian, which mimics measurement in standard paradigm, can amplify the probability of certain state to be ``measured", hence GSQC owns advantage over standard quantum computer.

A simple example, although of no practical interest, demonstrates this advantage: to teleport quantum state from qubit 1 to qubit 2, then to qubit 3, and so on to qubit $N$. By standard quantum computer, the probability to successfully realize this series of teleportations is $(1/4)^N$ because each teleportation process only has $1/4$ probability to succeed\cite{book}; while by GSQC, the probability is $(1-8/\lambda^2)^{3N}$: setting $\lambda=\sqrt{DN}$, then $P\approx e^{-24/D}$, and energy gap is $\Delta\propto \epsilon/(D^4N^4)$\cite{ours}. Thus GSQC only costs polynomially long time to finish the task while standard paradigm needs exponentially long time.

\begin{figure}
\begin{center}
\leavevmode
\hbox{\epsfxsize=6cm \epsffile{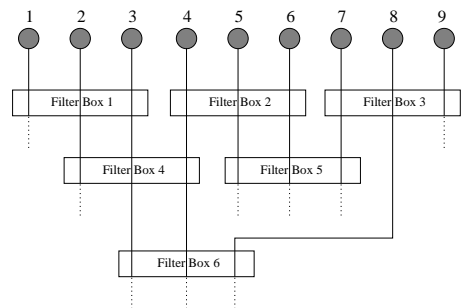}}
\end{center}
\caption{\small Design for circuit solving SAT problem with each clause involving 3 bits. Box labeled ``Filter Box" represents for circuit as shown in Fig.(\ref{filterbox}).}
\label{Circuit}
\label{circuit}
\end{figure}

The advantage of GSQC makes new quantum algorithm possible. Here I present a quantum algorithm to solve SAT problems as shown in Fig.(\ref{Circuit}),
a GSQC circuit to solve  a 3-SAT problem with only 9 bits. It's easy to extend to $N$-bit problem. Each clause is implemented by a ``filter box", and the circuit inside each filter box makes sure that on rows immediately below it the states satisfying clause $C_i$ have much larger amplitudes than other unsatisfying states, or we can say those unsatisfying states are filtered out. This can be realized by projection and boost Hamiltonians, and the detail will be given in the following example.
In the figure, the input state on the top row is $(|0\rangle+|1\rangle)(|0\rangle+|1\rangle)...(|0\rangle+|1\rangle)$, which is determined by the boundary Hamiltonian, $h^0=E(I-\sigma_x)$;
the clause involving qubit 1, 2 and 3 is implemented by filter box 1, the clause involving qubit 2, 3 and 4 implemented by filter box 4, the clause involving qubit 3, 4 and 8 implemented by filter box 6, etc.


When all constraints are implemented, at ground state the states measured on the final rows of the $N$ qubits should be superposition of states satisfying all constraints. I will show no backtracking is needed later.



Now I give an example on how to implement a filter box.
We  focus  on the 3-bit Exact Cover problem\cite{EC},  an instance of SAT problem, and belongs to NP-complete. Following is definition of 3-bit Exact Cover problem:

{\textit{There are $N$ bits $z_1,\ z_2,\ ...,\ z_N$, each taking the value 0 or 1. With $O(N)$   clauses  applied to them, each clause is a constraint involving three bits: one bit has value 1 while the other two have value 0. The task is to determine the $N$-bit state satisfying all the clauses. }}

\begin{figure}
\begin{center}
\leavevmode
\hbox{\epsfxsize=5cm \epsffile{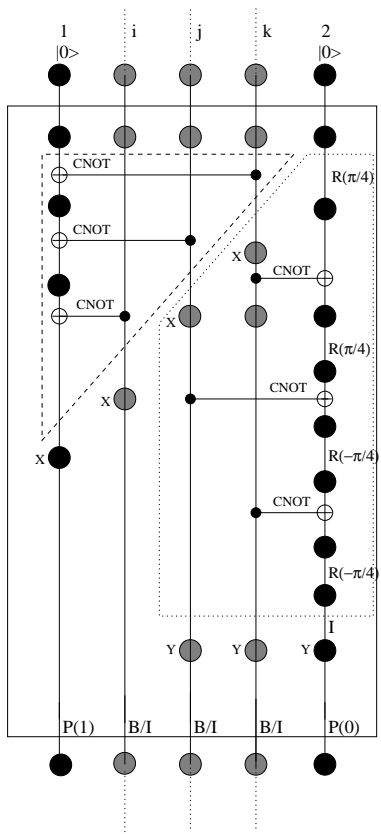}}
\end{center}
\caption{\small Design for "filter box". The labels  on the lines represent for corresponding Hamiltonians: $I$ for $h(I)$, $CNOT$ for $h(CNOT)$, $P(1)$ for projection $h(|1\rangle,\lambda)$ et. al. At the final rows, $B/I$ represents for boost Hamiltonian $h(B,\lambda)$ if there is no more clause to be applied to this qubit, otherwise, represents for identity Hamiltonian $h(I)$. There are teleportation boxes, not shown in figure, inserted on all qubits between two control Hamiltonians. Some dots marked by $X$ or $Y$ are for demonstration convenience in text.}
\label{filterbox}
\end{figure}
The algorithm is implemented by the circuit in Fig.(\ref{circuit}). Considering any one of the clauses, in GSQC  a filter box, involving three qubits $i,\ j$ and $ k$, which are represented by gray dot columns in Fig.(\ref{filterbox}), we add two ancilla qubits: qubit 1 and qubit 2, which are represented by dark dot columns.   Qubits $i,\ j,\ k$ at the first row are in the state $(|1\rangle+|0\rangle)$ if they have not experienced any clause yet, and the two ancilla qubits are in state $|\hat{0}\rangle$ and $|\tilde{0}\rangle$ on top rows by boundary Hamiltonians, where $ |\hat{\gamma}\rangle$ corresponds to state of ancilla qubit 1, and $ |\tilde{\gamma}\rangle$ to state of ancilla qubit 2. 

Inside the dashed triangle of Fig.(\ref{filterbox}),
after the first $CNOT$, we obtain state  $|\hat{1}{\rangle}|1{\rangle} + |\hat{0}{\rangle}|0{\rangle}$;
after the second $CNOT$:  $ |\hat{1}{\rangle}|1{\rangle}|0{\rangle}+|\hat{0}{\rangle}|0{\rangle}|0{\rangle} +  |\hat{0}{\rangle}|1{\rangle}|1\rangle+|\hat{1}{\rangle}|0{\rangle}|1{\rangle}$;
after the third $CNOT$: 
\eqa
&& |\hat{1}\rangle \left(|1\rangle |0\rangle |0\rangle+|0\rangle |1\rangle |0\rangle+|0\rangle |0\rangle |1\rangle+|1\rangle |1\rangle |1\rangle\right)\nonumber\\
&+&|\hat{0}\rangle \left(|1\rangle |1\rangle |0\rangle+|0\rangle |1\rangle |1\rangle+|1\rangle |0\rangle |1\rangle+|0\rangle |0\rangle |0\rangle\right).\nonumber
\eea
Immediately below the triangle, if the system stays at ground state, electron in ancilla qubit 1 is measured to be on the row labeled by $X$ and at state $|\hat{1}\rangle$, and the three electrons on qubit $i,\ j,\ k$ are all found on the rows labeled by $X$, then the three-qubit states satisfy the clause except for $ |1{\rangle}|1{\rangle}|1{\rangle}$.

The ancilla qubit $2$, starting at state $ |\tilde{0} {\rangle}$, experiences  $ CNOT$ gates controlled by qubits $j$ and $k$, and $R(\pm \pi/4)$ is defined in \cite{Tof} as $R_y(\pm \pi/4)$, as shown within the dotted pentagon in Fig.(\ref{filterbox}). All those transformations happened  inside the dotted pentagon are equivalent to a Toffoli gate except for some unimportant phases\cite{Tof}: if both qubits $j$ and $k$ are in state $|1\rangle$, then the ancilla qubit $2$ reverses to state $|\tilde{1}\rangle$, otherwise, it remains at state $|\tilde{0}\rangle$. After this nearly Toffoli transformation, if at ground state electrons in qubit $j,\ k$ and ancilla qubit $2$ are found on rows labeled by $Y$, and ancilla qubit 2 is at $|\tilde{0}\rangle$, then the three qubits will be at $|\tilde{0} \rangle ( |0 \rangle |0\rangle + |1\rangle |0\rangle + |0\rangle |1\rangle)$.
 Thus if at ground state all electrons are found on rows immediately below both the dashed triangle and the dotted pentagon, and if ancilla qubit 1 is at $|\hat{1}\rangle$ and ancilla qubit 2 at $|\tilde{0}\rangle$, then  the three qubits $i,\ j,\ k$ and two ancilla qubits will  satisfy the clause:
\eqa
|\hat{1}\rangle|\tilde{0}\rangle\left(|1\rangle|0\rangle|0\rangle+|0\rangle|1\rangle|0\rangle+|0\rangle|0\rangle|1\rangle
\right).
\label{ff}
\eea

In order to make the right states pass through the filter box with large probability, we add projection Hamiltonians and boost Hamiltonians as shown in the lower part of Fig.(\ref{filterbox}).
The projection Hamiltonians on final rows of two ancilla qubits limit and amplify them at the states we prefer:   ancilla qubit 1 at $|\hat{1}\rangle$, and   ancilla qubit 2 at $|\tilde{0}\rangle$.
 If a qubit doesn't experience any more clause, it will end with a boost Hamiltonian, otherwise, its quantum state will be teleported to a new qubit through teleportation box, not shown in Fig.(\ref{filterbox}), and the new qubit continues experience more clauses. Thus the projection Hamiltonian on two ancilla qubits and boost Hamiltonian on the three qubits make sure that the ground state wavefunction concentrated on the final rows in Fig.(\ref{filterbox}) with state at Eq.(\ref{ff}).

 Noting that in the filter box all the three qubits $i,\ j,$ and $k$ always act as control qubits, thus the entanglement of these three qubits with other qubits not involved in this particular clause still keep the same. When adding a clause, the resulted states satisfying this clause will also satisfy all previous applied clauses. Thus unlike classical algorithm, no backtracking is needed.

In the circuit of Fig.(\ref{circuit}), if there is at least one  solution, and all electrons are simultaneously found on the final rows of all qubits, then the reading of the $N$-bit states satisfying all constraints.

In order to keep the energy gap from being too small, like in \cite{ours}, on every qubit teleportation boxes are inserted between two control Hamiltonians, thus the total number of qubits increases while the energy gap $\Delta\propto\epsilon/\lambda^8$ if in all boost and projection Hamiltonian the amplifying factors have the same value $\lambda$.

 For one clause, or a filter box,  it needs 10 teleportation boxes (each teleportation box adds two more qubits) on the original five-qubit circuit, noting that  on the end of qubit $i,\ j$ and $k$ in Fig.(\ref{filterbox}) teleportation boxes are needed because more clause will be added. Thus adding one more filter box means adding 20 more qubits.
 The  number of clause for a NP hard 3-bit Exact Cover problem is about the same order as the number of bits $N$\cite{3SAT}, say $ \alpha N$ with $\alpha$ being $O(1)$, then there are totally $20\alpha N$ qubits and each of them ends with either projection or boost Hamiltonian. Probability of finding all electrons at the final rows is approximately
\eqa
P\approx \left(1-{C}/{\lambda^2}\right)^{20\alpha N},
\eea
where $C=8$, the length of the longest qubit\cite{ours}. It is assumed that, at ground state, in each filter box the ancilla qubit 1 and 2 have appreciable probability in $|1\rangle$ and $|0\rangle$ states respectively before projection Hamiltonians. We will address situation when the assumption is violated.

In order to make the probability independent of number of bits $N$, we take $\lambda^2=DN$, where $D$ is an arbitrary number. Then as $N$ becomes large, we obtain
\eqa
P\approx\left(1-{C}/{(DN)}\right)^{20\alpha N}\approx e^{-20\alpha C/D},
\label{P}
\eea
and energy gap\cite{ours}
\eqa
\Delta\propto \epsilon/\lambda^8\propto {\epsilon}/{(D^4N^4)},
\eea
from which one can estimate time cost.

To make the GSQC circuit at ground state,
 we can use adiabatic approach: first we set $\lambda=1$ for boost and projection Hamiltonian on final rows of all qubits, and replace the single qubit Hamiltonian between the first two rows of all qubits by a boost Hamiltonian 
\eqa
h'(B,\lambda')=\epsilon\left[
\frac{1}{\lambda'^2}  C^{\dagger}_{1}C_{1}+  C^{\dagger}_{2}C_{2}
-\frac{1}{\lambda'}\left(C^{\dagger}_{1}C_{2}+h.c.\right)\right],
\eea
so that the wavefunction amplitude of the first row is boosted as $\lambda'\gg 1$. Now in the ground state the electrons concentrate at the first rows as $1/\lambda'\rightarrow 0$, thus the ground state is easy to be prepared, and energy gap $\Delta\propto\epsilon/n^2$ with $n= 8$ being the length of longest qubit. The next step is turning the quantity $1/\lambda'$ to 1 adiabatically, during which the energy gap remains at $\epsilon/n^2$ and the ground state wavefunction spreads to other rows from the first row.  The third step is turning $1/\lambda$ from 1 to $1/\sqrt{DN}$ adiabatically. In this process the energy gap   decreases monotonically from $\epsilon/n^2$ to what we obtained above: $\epsilon/D^4N^4$, and wavefunction concentrated on the final rows of all qubit as we wish. Thus the scale of time cost is about $T\propto 1/\Delta^2\propto N^8$\cite{Farhi0}, local adiabatic approach may reduce the time cost further\cite{local}.

Above analysis is under the assumption that the number of satisfying states gradually decreases as the clauses are implemented one by one. There is a situation that might hurt our algorithm: 
after adding one more clause, if the number of satisfying states drops dramatically, our algorithm will be hurt. For example, if one constructs GSQC for the Grover's search problem with one condition to find a unique satisfying state from $2^N$ states, then he will find that there is an ancilla qubit containing such unnormalized state 
\eqa
|0\rangle|\text{satisfying}\rangle+\sum_{i=1}^{2^{N}-1}|1\rangle|\text{unsatisfying}^{(i)}\rangle
\eea
before the projection Hamiltonian $h(|0\rangle,\lambda)$. In order to amplify the amplitude of the correct state on the final row, it requires $\lambda\ge 2^{N/2}$, which makes the energy gap exponentially small.

 Does this happen to general SAT problems? In \cite{nature} it was suggested that close to threshold $\alpha_c$   computational complexity might be related with the forming of backbone, each of a subset of bits has average value close to 1 or 0 in the subspace of satisfying states. The existence of backbone means that most satisfying states contain the state represented by backbone, and if adding one more clause kicks out the states consistent with backbone from satisfying subspace, the number of satisfying states drops dramatically. With advantage over classical algorithm, performance of  our algorithm is not affected by forming of backbone, however, as more clauses applied, the disappearance of already existed backbone in the satisfying subspace might hurt.  There is a criterion determining efficiency of our algorithm: the ratio $S_j/S_{j+1}$, with $S_j$ being the number of solutions when the $j$th clause is applied, and $S_{j+1}$ the number of solutions when the $(j+1)$th clause is applied. For example, $S_0/S_1=8/3$ for 3-bit Exact Cover problem.
 If $S_j/S_{j+1}\gg 1$, on the ancilla qubit of the $(j+1)$th filter box, the probability of finding electron on its final row will be $p\approx(1-C S_j/(\lambda^2 S_{j+1}))$. In order to have appreciable probability as Eq.(\ref{P}), it requires $\lambda^2$ increase from $DN$ to $DNS_j/ S_{j+1}$, hence the energy gap is also suppressed. In advance one cannot know what value $S_j/S_{j+1}$ is, thus a overhaul factor for $\lambda$ is needed.
 If this ratio happens to be exponentially large, then our algorithm cannot solve the SAT problem in polynomial time. However, one might be able to identify backbone by trials, and then choose proper order to implement clauses so that $S_j/S_{j+1}$ can be kept small.

In conclusion, we have demonstrated that a ground state quantum computer can solve a general SAT problem. A specific example, the 3-bit Exact Cover problem, is given.
We show that an 3-bit Exact Cover problem can be solved by the quantum algorithm described here, and  the time cost is related with the number of bits $N$ and parameter $S_j/S_{j+1}$. If $S_j/S_{j+1}$ stays small or only polynomially large, then the presented algorithm can solve this SAT problem in polynomial time.

I would like to thank A. Mizel for helpful discussion. This
work was supported in part by the NSF under grant \# 0121428 and
by ARDA and DOD under the DURINT grant \# F49620-01-1-0439.

\end{document}